\documentclass[useAMS,usenatbib]{mn2e}
\usepackage{natbib}
\usepackage{color,graphicx} 
\usepackage{amsmath}        
\usepackage{amsfonts}       
\usepackage{amssymb}        
\title[A search for optical bursts from RRAT J1819$-$1458: II.]{A search for optical bursts from RRAT J1819$-$1458: II. Simultaneous ULTRACAM--Lovell Telescope observations}
\author[V. S. Dhillon et al.]{V. S. Dhillon,$^{1}$\thanks{E-mail: vik.dhillon@sheffield.ac.uk} 
E. F.~Keane,$^{2}$ T. R. Marsh,$^{3}$ B. W. Stappers$^{2}$
\newauthor
C. M. Copperwheat,$^{3}$ R. D. G. Hickman,$^{3}$ C. A. Jordan,$^{2}$ P. Kerry,$^{1}$ 
\newauthor
M. Kramer,$^{2,4}$ S. P. Littlefair,$^{1}$ A. G. Lyne,$^{2}$ R. P. Mignani,$^{5,6}$  A. Shearer$^7$ \\
$^{1}$Department of Physics and Astronomy, University of Sheffield, Sheffield S3 7RH, UK \\
$^{2}$Jodrell Bank Centre for Astrophysics, School of Physics and Astronomy, University of Manchester, Manchester M13 9PL, UK \\
$^{3}$Department of Physics, University of Warwick, Coventry CV4 7AL, UK \\
$^{4}$Max Planck Institut f\"{u}r Radioastronomie, Auf dem H\"{u}gel 69, 53121 Bonn, Germany \\
$^{5}$Mullard Space Science Laboratory, University College London, Holmbury St. Mary, Dorking, Surrey, RH5 6NT, UK \\
$^{6}$Institute of Astronomy, University of Zielona G\'{o}ra, Lubuska 2, 65-265 Zielona G\'{o}ra, Poland \\
$^{7}$Centre for Astronomy, National University of Ireland, Galway, Newcastle Rd., Galway, Ireland \\}

\begin{document}

\date{Accepted for publication in the Monthly Notices of the Royal
  Astronomical Society on 2011 March 7}

\maketitle

\begin{abstract}
  The Rotating RAdio Transient (RRAT) J1819$-$1458 exhibits $\sim
  3$\,ms bursts in the radio every $\sim 3$\,min, implying that it is
  visible for only $\sim 1$\,s per day. Assuming that the optical
  light behaves in a similar manner, long exposures of the field would
  be relatively insensitive due to the accumulation of sky photons. A
  much better way of detecting optical emission from J1819$-$1458
  would then be to observe with a high-speed optical camera
  simultaneously with radio observations, and co-add only those
  optical frames coincident with the dispersion-corrected radio
  bursts. We present the results of such a search, using simultaneous
  ULTRACAM and Lovell Telescope observations. We find no evidence for
  optical bursts in J1819$-$1458 at magnitudes brighter than $i'=19.3$
  (5$\sigma$ limit). This is nearly 3 magnitudes fainter than the
  previous burst limit, which had no simultaneous radio observations.
\end{abstract}


\section{Introduction}
The RRATs are a group of isolated Galactic neutron stars discovered in
an archival search of the Parkes Multi-beam Pulsar Survey (PMPS;
\citealt{mclaughlin06}). They are characterised by millisecond radio
bursts with flux densities at $1.4$~GHz of $\sim0.1-10$~Jy. The bursts
are infrequent, with intervals of as long as a few hours to as short
as a few minutes~\citep{keane10}. The rotation periods, inferred by
dividing the burst intervals by the largest common denominator, lie in
the $0.1-7$~s range. Of the 22 RRAT sources in the PMPS with known
periods, 11 have periods greater than 4~s, compared to just $\sim2\%$
of radio pulsars~\citep{keane10a}. These long periods are reminiscent
of the (X-ray dim) isolated neutron stars and the magnetars (see
\citealt{haberl07} and \citealt{mereghetti08}). Using their dispersion
measure, the RRATs are found to lie approximately 1--7 kpc
distant in the Galactic plane~\citep{keane10a}.

The nature of RRATs is still an open question.  A simple estimate of
their Galactic population suggests that they may be more abundant than
the radio pulsars. In fact, if we consider RRATs and the other known
classes of neutron stars as distinct populations, a birthrate problem
arises, i.e. neutron stars appear to be formed faster than the
observed supernova rate. This problem may be resolved, however, if the
various neutron star types are instead regarded as different
evolutionary phases~\citep{keane08}. Two main models have been
proposed for the intermittent pulses exhibited by RRATs, with some
researchers assigning the phenomenon to detection issues and others
favouring intrinsically transient emission. \citet{Weltevrede06}
suggest that RRATs are distant analogues of PSR B0656+14,
i.e. pulsars with regular emission that is below our detection limit
but which show large amplitude pulses drawn from an extended
pulse-energy distribution.  Alternatively, \citet{cordes08} suggest
that the bursts in RRATs are powered by the episodic injection of
material from a circumpolar asteroid belt, formed from supernova
fallback material, that temporarily reactivates a quiescent region of
the magnetosphere. Similarly, \citet{luo07} propose that RRATs may be
surrounded by planetary-like radiation belts, and the trapped plasma
in this belt is episodically disrupted (e.g. by starquakes or stellar
oscillations) causing particle precipitation towards the star and
hence bursts. Besides external triggers, it has also been
suggested~\citep{zhang07} that RRATs are old pulsars approaching the
`death valley'~\citep{cr93}, where pulsar emission is thought to
switch off.

Of the 56 known RRATs~\citep{keane10a}, J1819$-$1458 is the most
prolific, brightest and thus best-studied source. J1819$-$1458 has a
rotation period of $4.263$~s and shows $\sim3$~ms pulses every
$\sim3$\,min, amounting to $\sim1$~s of radio emission per day. From
its measured period derivative, J1819$-$1458 has an inferred magnetic
field strength of $B=5\times10^{13}$~G, just below that of the
magnetars, providing another link between these two classes of neutron
stars. As well as in the radio (e.g. \citealt{lyne09}), J1819$-$1458
has been observed in X-rays on several occasions
(\citealt{reynolds06}, \citealt{mclaughlin07}, \citealt{rea09}). The
X-ray observations show a thermal spectrum consistent with what is
expected from a cooling neutron star, and the X-ray flux exhibits the
same rotation period as derived from radio observations. Observations
of J1819$-$1458 at wavelengths other than radio and X-ray are also
highly desirable in order to measure the spectral energy distribution
and help constrain the pulsar emission mechanism. Deep infrared
observations have revealed very tentative evidence for a counterpart
at $K_{\mathrm{s}}\sim 21$ magnitude \citep{rea10}. There is no
evidence of an optical counterpart, but this could be due to the
rather modest magnitude limit of $I=17.5$ \citep{reynolds06}.

Taking longer exposures to go deeper, however, is not necessarily the
best solution, as the RRATs may have very faint persistent optical/IR
emission and only emit strongly at these wavelengths during
bursts\footnote{The main optical pulse of the Crab pulsar, for
  example, is $\sim 5$ magnitudes brighter than its persistent light
  level.}. In this case, the best strategy would be to reduce the
contribution of the sky and take a continuous sequence of extremely
short exposures on a large-aperture telescope covering a number of
burst cycles in order to catch a burst in one or two of the frames. In
paper I \citep{dhillon06}, we tried such an approach using the
high-speed CCD camera ULTRACAM \citep{dhillon07b} on the 4.2-m William
Herschel Telescope (WHT). We found no evidence for bursts brighter
than $i'=16.6$. This limit may not appear to be particularly deep, but
it must be remembered that it refers to the burst magnitude, not the
persistent magnitude. In fact, there is only one way in which it is
possible to significantly improve upon the ULTRACAM+WHT burst limit:
observing in the optical simultaneously with the radio, which would
allow just those optical frames coincident with the
dispersion-corrected radio bursts to be searched for optical
bursts. In this paper we report on such observations, obtained with
ULTRACAM on the WHT and the 3.5-m New Technology Telescope (NTT),
simultaneously with the 76-m Lovell Telescope at Jodrell Bank
Observatory (JBO).

\section{Observations}
\label{sec:observations}
\begin{figure*}
  \centering
  \includegraphics[trim = 80mm 10mm 60mm 20mm, clip, width=4.5cm,angle=270]{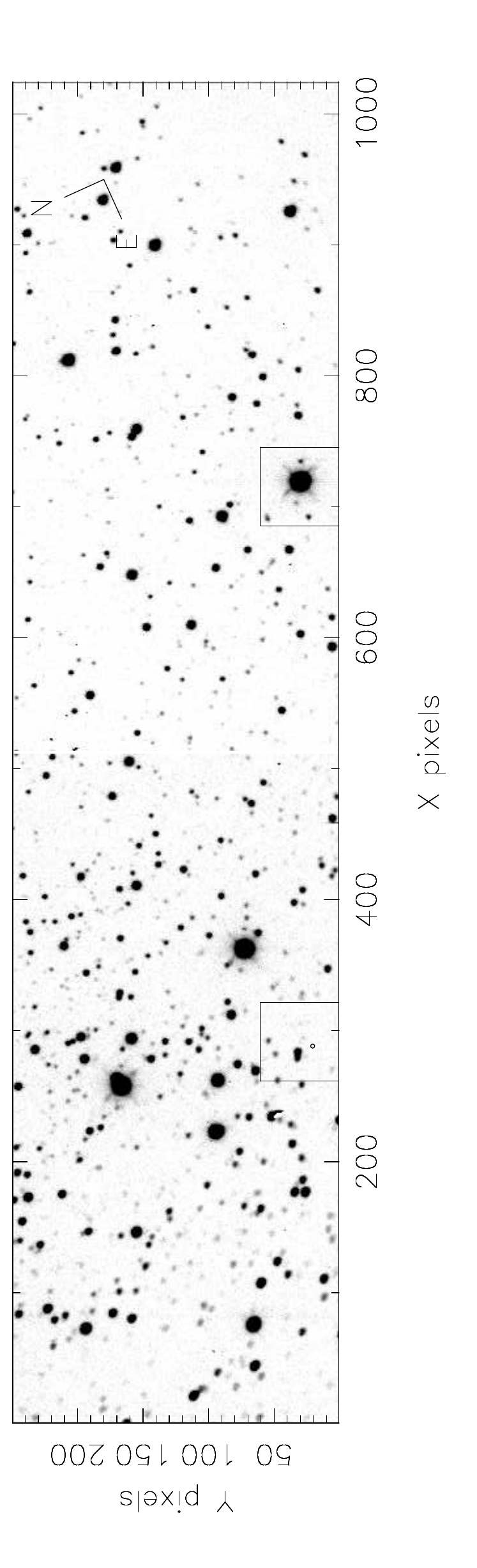}
  \includegraphics[trim = 25mm 10mm 10mm 15mm, clip, scale=0.25,angle=270]{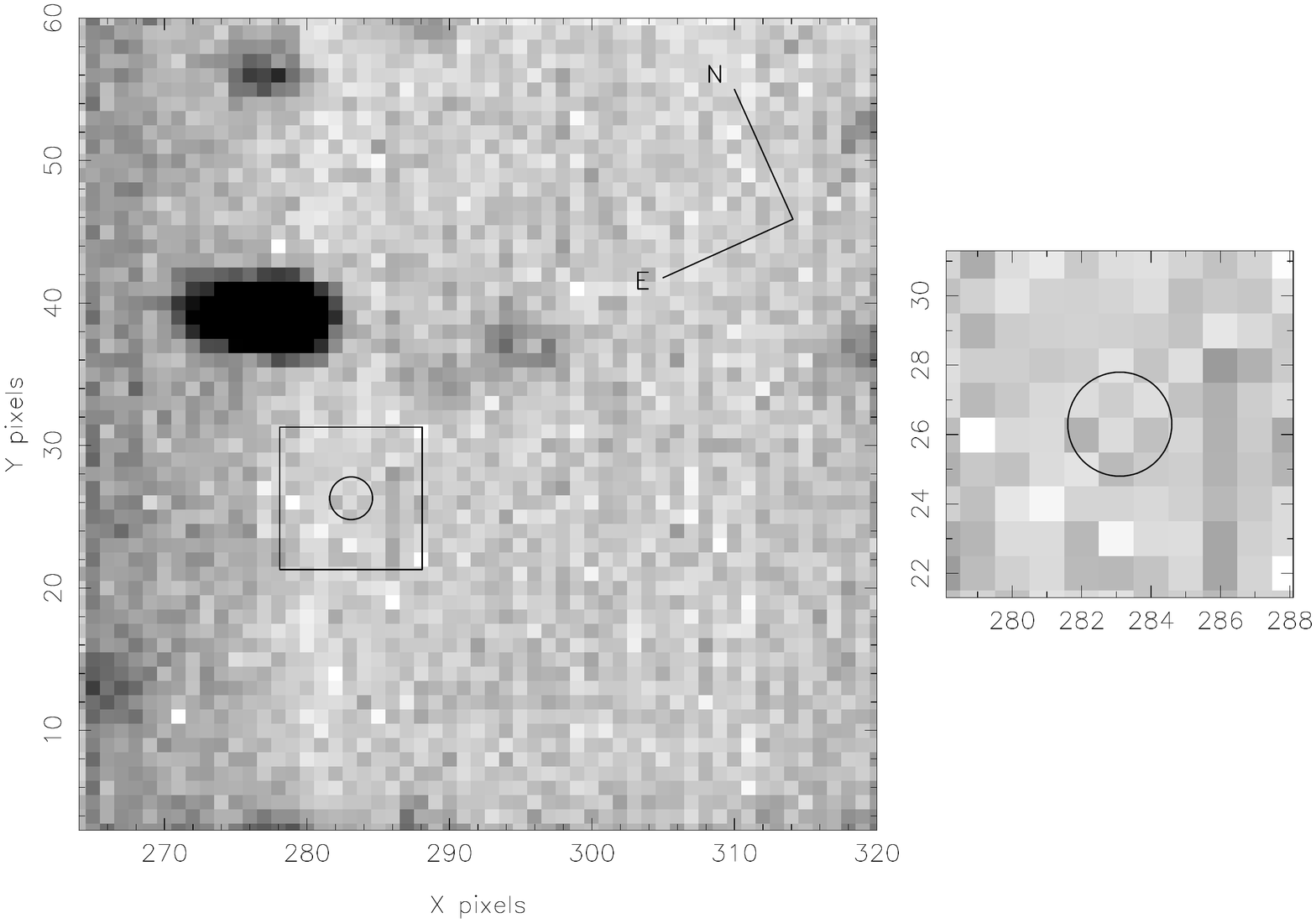}~\includegraphics[trim = 25mm 10mm 10mm 15mm, clip, scale=0.25,angle=270]{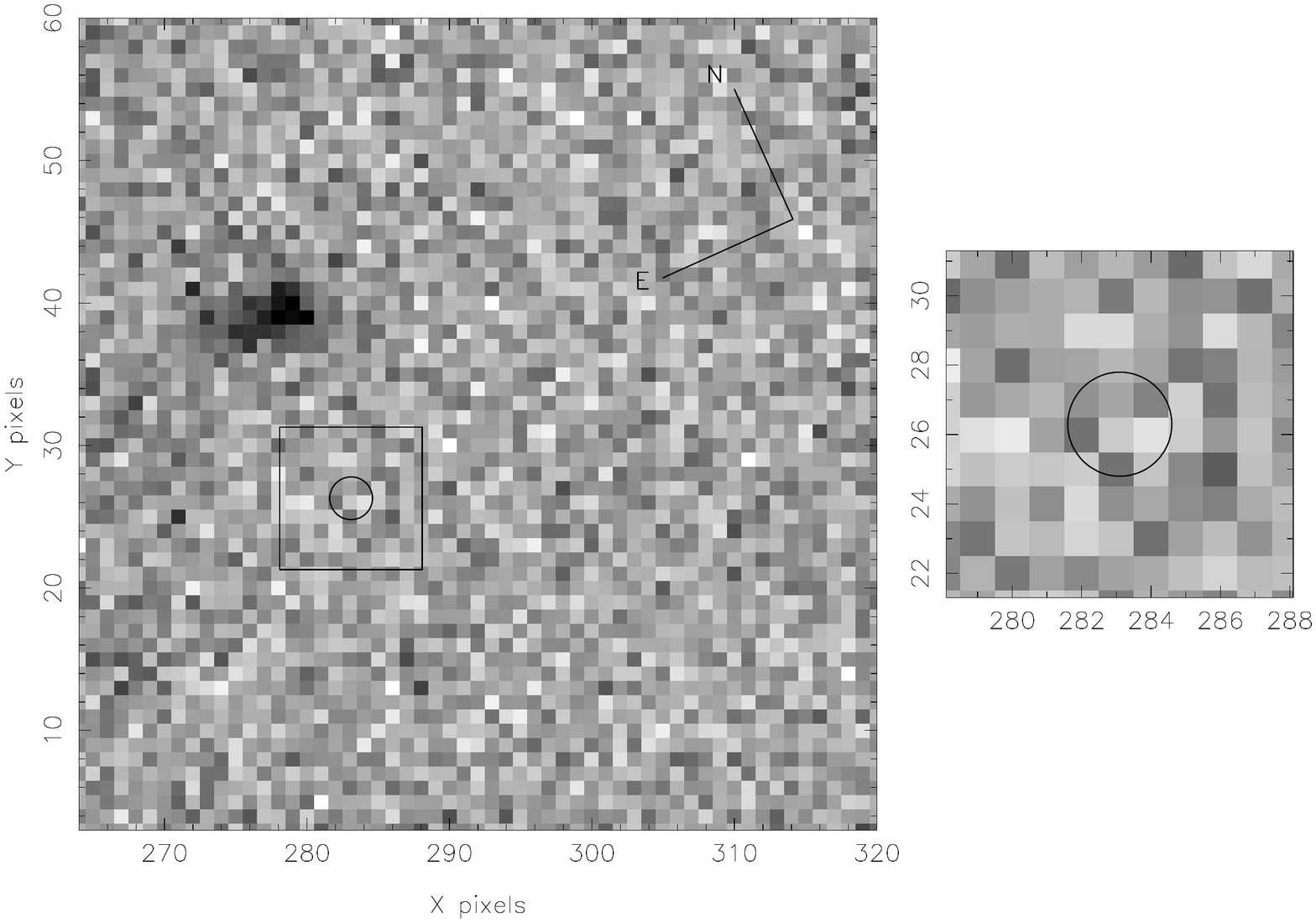}
  \includegraphics[trim = 11mm 0mm 14mm 15mm, clip, scale=0.25,angle=0]{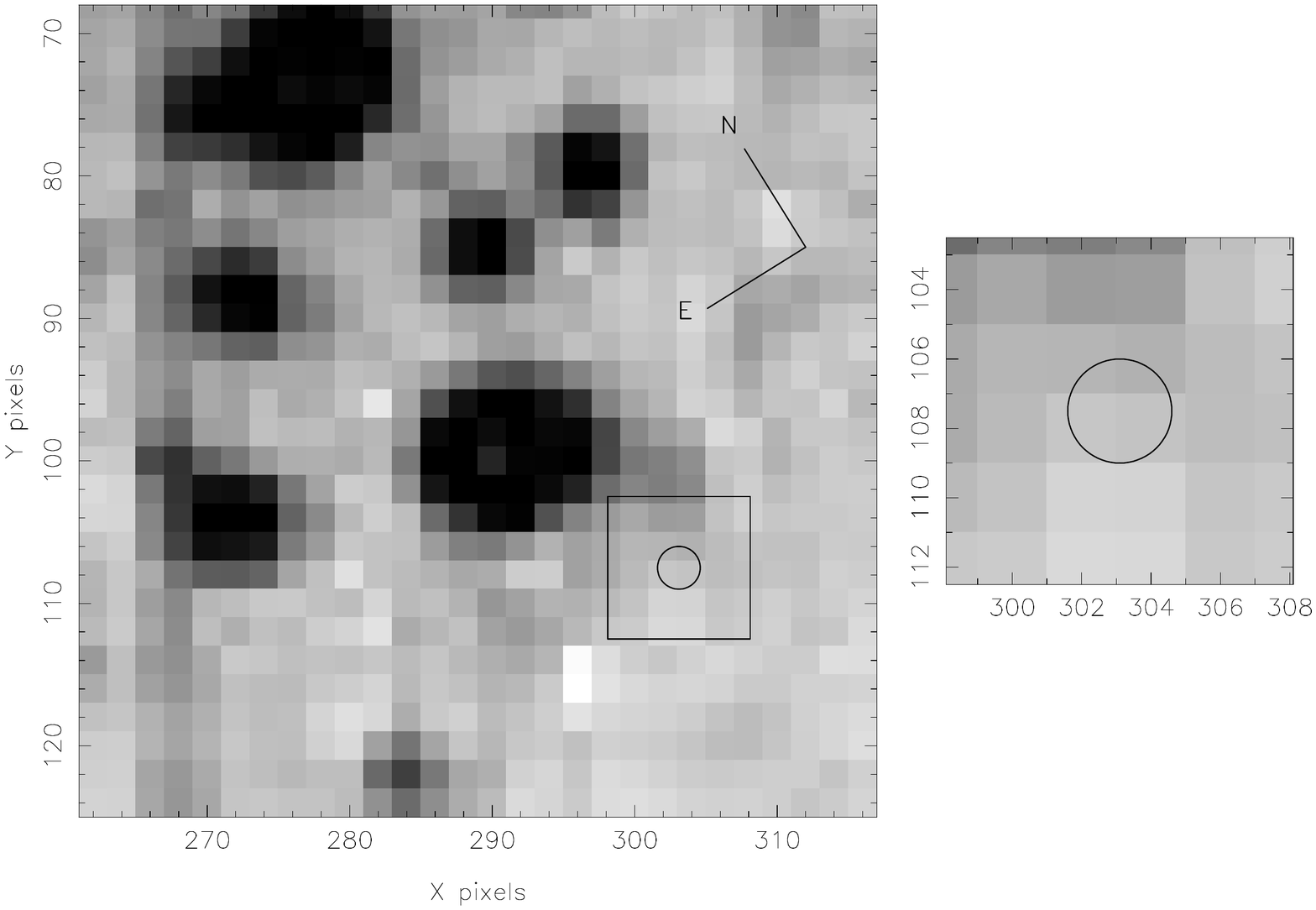}~  \includegraphics[trim = 11mm 0mm 10mm 15mm, clip, scale=0.25,angle=0]{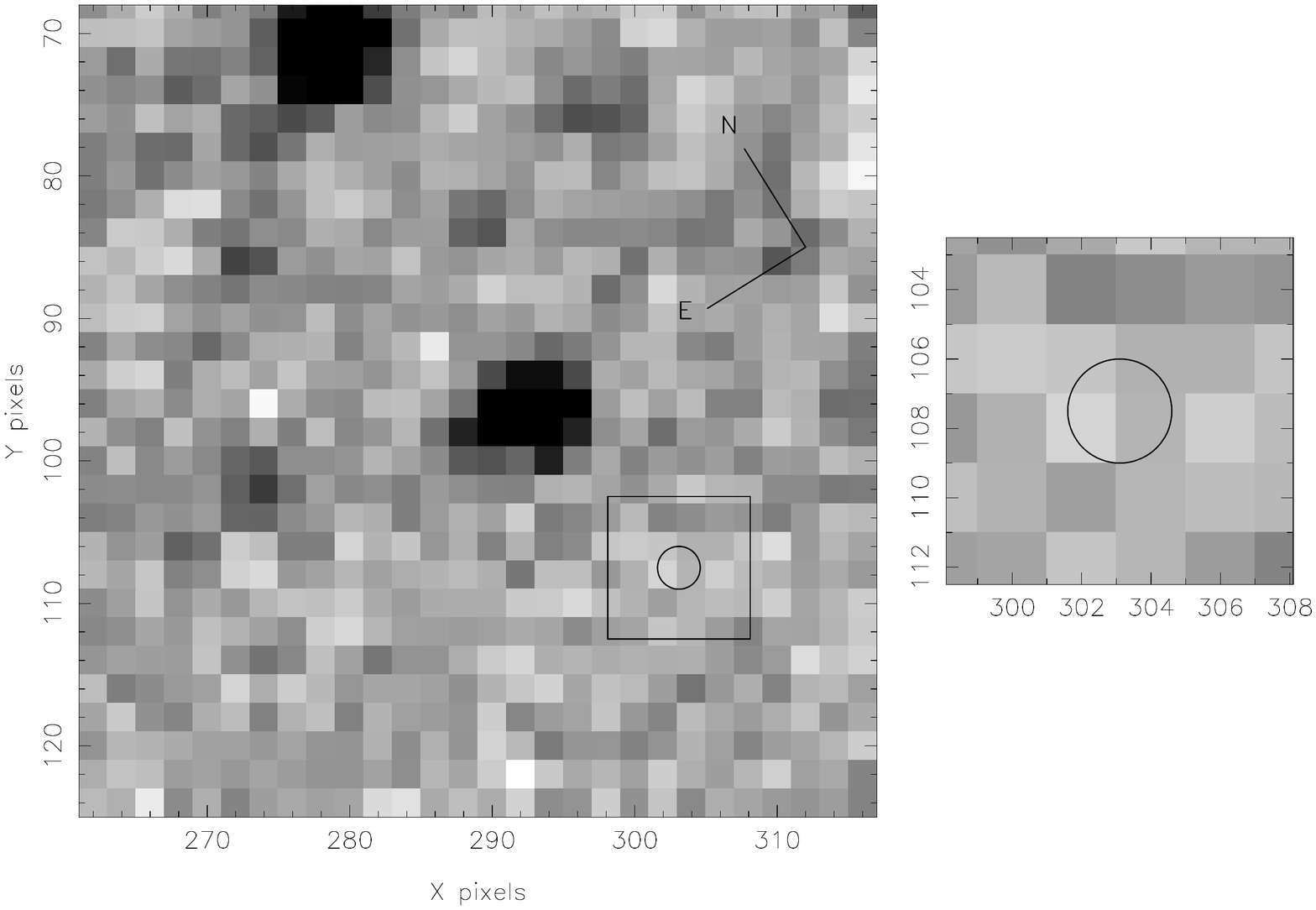}
  \caption{Top panel: WHT+ULTRACAM image of the field of J1819$-$1458
    in the $i'$-band, taken by summing 40 acquisition frames on 2008
    August 6 with a total exposure time of 127\,s.  The two ULTRACAM
    drift-mode windows used to acquire high-speed data on J1819$-$1458
    and the comparison star are shown by the boxes. The X-ray position
    of J1819$-$1458 derived by \protect\cite{rea09} is marked as a dot
    just lower-left of centre in the left-hand window. The plate scale
    is 0.3 ''/pixel and the orientation of the field is shown at the
    upper right. Middle panel, left: Summed image of the J1819$-$1458
    window on 2008 August 6, containing all 112\,588 WHT+ULTRACAM
    frames obtained on that night. The small square shows the area
    plotted at a larger scale to the right. The circle has a radius of
    0.5'' and is centred on the X-ray position of the RRAT. The plate
    scale is 0.3 ''/pixel and the orientation of the field is shown at
    the upper right. Middle panel, right: As for the left, but showing
    only the sum of the 24 WHT+ULTRACAM frames coincident with
    dispersion-corrected radio bursts on 2008 August 6. Bottom panel, left: As for the
    middle-left panel, but showing the sum of all 68\,274 NTT+ULTRACAM
    frames obtained on 2010 June 14. Note that each pixel is 0.7''
    (after binning) and the orientation of the field is slightly
    different compared to 2008 August 6. Bottom panel, right: As for
    the left, but showing only the sum of the 25 NTT+ULTRACAM frames
    coincident with radio bursts on 2010 June 14.}
  \label{fig:finding_charts}
\end{figure*}

The observations of J1819$-$1458 were obtained on the nights of 2008
August 6 (WHT+ULTRACAM and JBO) and 2010 June 14 (NTT+ULTRACAM and
JBO). In addition to shot noise from any object flux, every ULTRACAM
data frame has noise contributions from the sky and CCD readout
noise. The sky noise can be reduced by reducing the exposure time, but
the readout noise cannot. Hence it makes sense to expose each data
frame for as long as the readout noise is the dominant noise source,
thereby maximising the chances of observing a burst in a single frame
without significantly degrading the signal-to-noise ratio of the data.
ULTRACAM was hence used in drift mode, which gives the highest frame
rate (see \citealt{dhillon07b}), with one window centred on the X-ray
position of the RRAT \citep{rea09} and the other on a nearby
comparison star, as shown in the top panel of
Fig.~\ref{fig:finding_charts}. An SDSS $i'$ filter and the slow
readout speed were used in the red arm of ULTRACAM on both nights, and
the focal-plane mask was used to prevent light from bright stars and
the sky from contaminating the windows (see \citealt{dhillon07b}). On
2008 August 6, the CCD windows were unbinned and of size $60\times60$
pixels, where each pixel on the WHT is 0.3''. A total of 112\,588
frames were obtained between 21:11--22:49 UTC on this night, each of
51.1\,ms exposure time and 1.4\,ms dead time. The data were obtained
in photometric conditions, with no Moon and seeing of 0.9''. On 2010
June 14, the CCD windows were binned $2\times2$ and of size
$150\times150$ pixels, where each unbinned pixel on the NTT is
0.35''. A total of 68\,274 frames were obtained between 01:32--02:14
UTC and 02:45--03:46 UTC; the gap in the middle of the run was due to
a GRB override observation~\citep{dhillon10}. Each frame had an
exposure time of 86.5\,ms and a dead time of 3.5\,ms. Conditions on
this night were not as good as in 2008, with seeing of 1.9'' at the
start of the run, dropping to 1.2'' at the end. The night was
photometric and there was no Moon.

Simultaneous radio observations at JBO were made at an observing
frequency of $1.4$~GHz using a dual-channel cryogenic receiver
sensitive to left- and right-handed circular polarisation. In 2008 an
analogue filterbank (AFB) backend was used with an observing bandwidth
consisting of $64\times1$~MHz channels and a time sampling of
$100$~$\mu$s. Since 2009, pulsar observations at JBO have upgraded to
using a digital filterbank (DFB) backend. Thus the 2010 observations
used the DFB with a bandwidth of $1024\times0.5$~MHz channels, half of
which ($\sim250$ MHz) were usable, and a time sampling of $1$~ms. In
both cases, the polarisations were summed to give total intensity
(Stokes I) and the output was either 1-bit digitised (in 2008) or
2-bit digitised (in 2010). In both radio datasets the zero dispersion-measure
subtraction algorithm~\citep{eatough09} was used in an attempt to
remove sources of broadband radio frequency interference, as described
by \citet{keane10}. The 2008 dataset suffers less from radio frequency
interference than the 2010 observations due to the narrower bandwidth
of the former. The pulse times of arrival at JBO were then obtained by
cross-correlating the single-pulse profiles with a smooth
single-component template. In addition to the light travel time from
the source to the Earth, radio signals traversing the interstellar medium
suffer an extra frequency-dependent delay of the form:
$t_{\mathrm{DM}}=4150\;\mathrm{s}\frac{DM}{f^2}$, where $f$ is the
observing frequency in MHz, and $DM$ is the dispersion measure (the
integrated electron density along the line of sight to the source
measured by convention in units of
$\mathrm{cm}^{-3}\,\mathrm{pc}$). At $1.4$ GHz, this delay is $415$~ms
for J1819$-$1458 ($DM=196.0(4)\;\mathrm{cm}^{-3}\,\mathrm{pc}$) with
respect to a signal at infinite frequency, and differs by $38$~ms
($250$~ms) between the top and bottom of our band for the 2008 (2010)
observation. The optical signal is not subject to such a delay. Thus
the radio signal is de-dispersed to infinite frequency (i.e. this
delay is removed according to the known dispersion measure of the
source) before comparison is made with the optical~times.

\section{Data reduction}
The ULTRACAM frames were first debiased and then flat-fielded using
images of the twilight sky.  The list of JBO pulse arrival times on
each night were corrected for dispersion, converted to Barycentric
Dynamical Time (TDB) at the solar system barycentre, and then compared
with the barycentred ULTRACAM times. Note that each ULTRACAM frame is
time-stamped to a relative (i.e. frame-to-frame) accuracy of
$\sim$\,50\,$\mu$s and an absolute accuracy of $\sim$\,1 ms using a
dedicated GPS system (see \citealt{dhillon07b}). It was found that 24
and 25 dispersion-corrected radio bursts had corresponding optical
frames on 2008 August 6 and 2010 June 14, respectively, and these
ULTRACAM images were then shifted to correct for telescope guiding
errors and co-added (see Section~\ref{sec:results}).

Aperture photometry at the X-ray position of
J1819$-$1458~\citep{rea09} was performed using the ULTRACAM pipeline
data reduction system. To do this, we had to determine the pixel
position of J1819$-$1458 on the ULTRACAM CCD. This was achieved by
transforming the $x,y$ pixel coordinates to equatorial coordinates
using the known positions of bright stars in the field. We estimate
that the uncertainty in the resulting pixel position of J1819$-$1458
is 0.5''. We extracted a light curve for both the comparison star and
the position of the RRAT using variable-sized apertures with diameters
set to 3 times the seeing, as measured from the FWHM of the comparison
star, which is $\sim 3-6$ times larger than the error in the RRAT
position on the ULTRACAM frames.  The sky level was determined from an
annulus surrounding each aperture and subtracted from the object
counts.

\begin{figure*}
  \centering
  \includegraphics[trim = 15mm 10mm 40mm 5mm, clip, scale=0.32,angle=0]{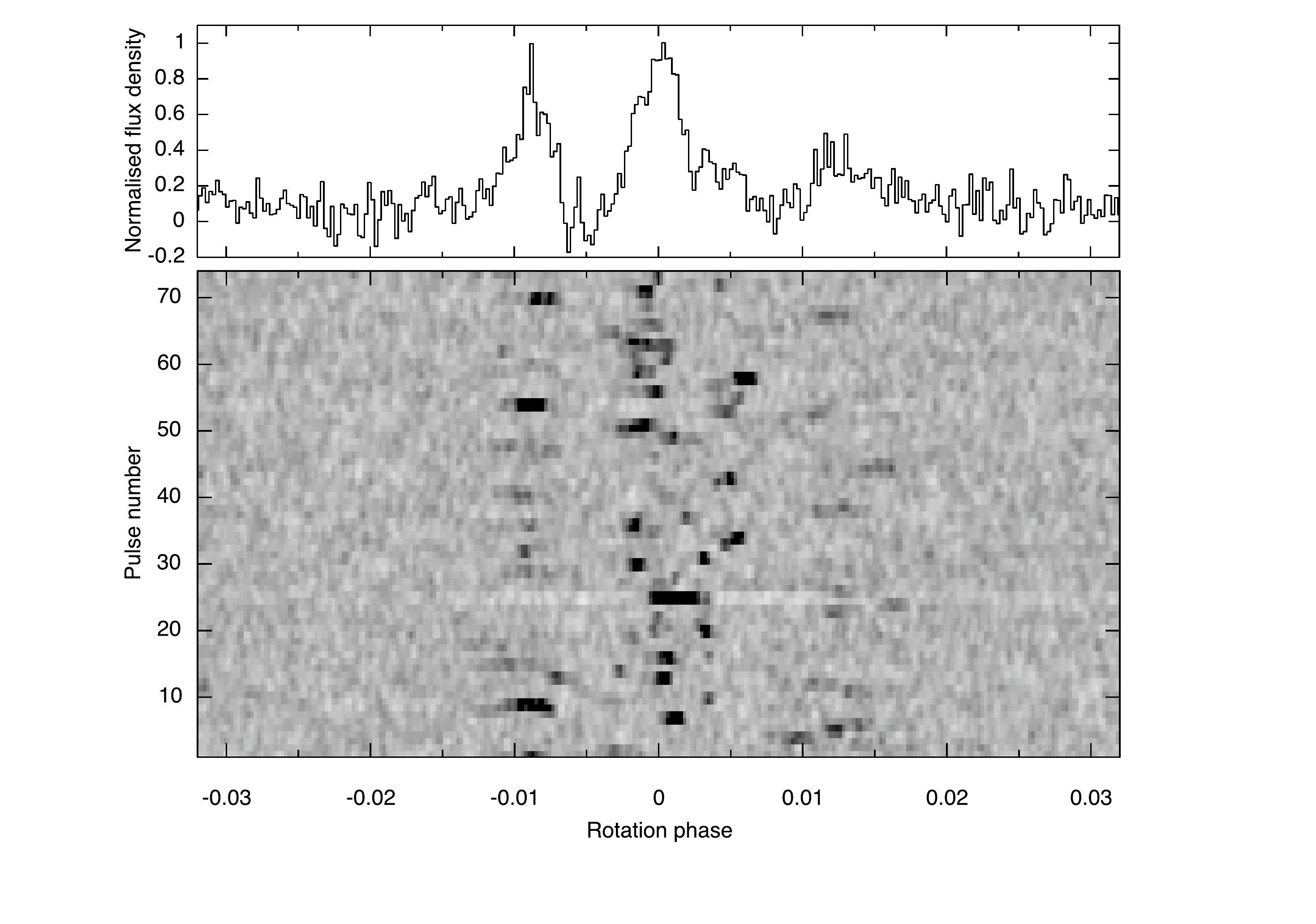}\includegraphics[trim = 15mm 10mm 5mm 5mm, clip, scale=0.34,angle=0]{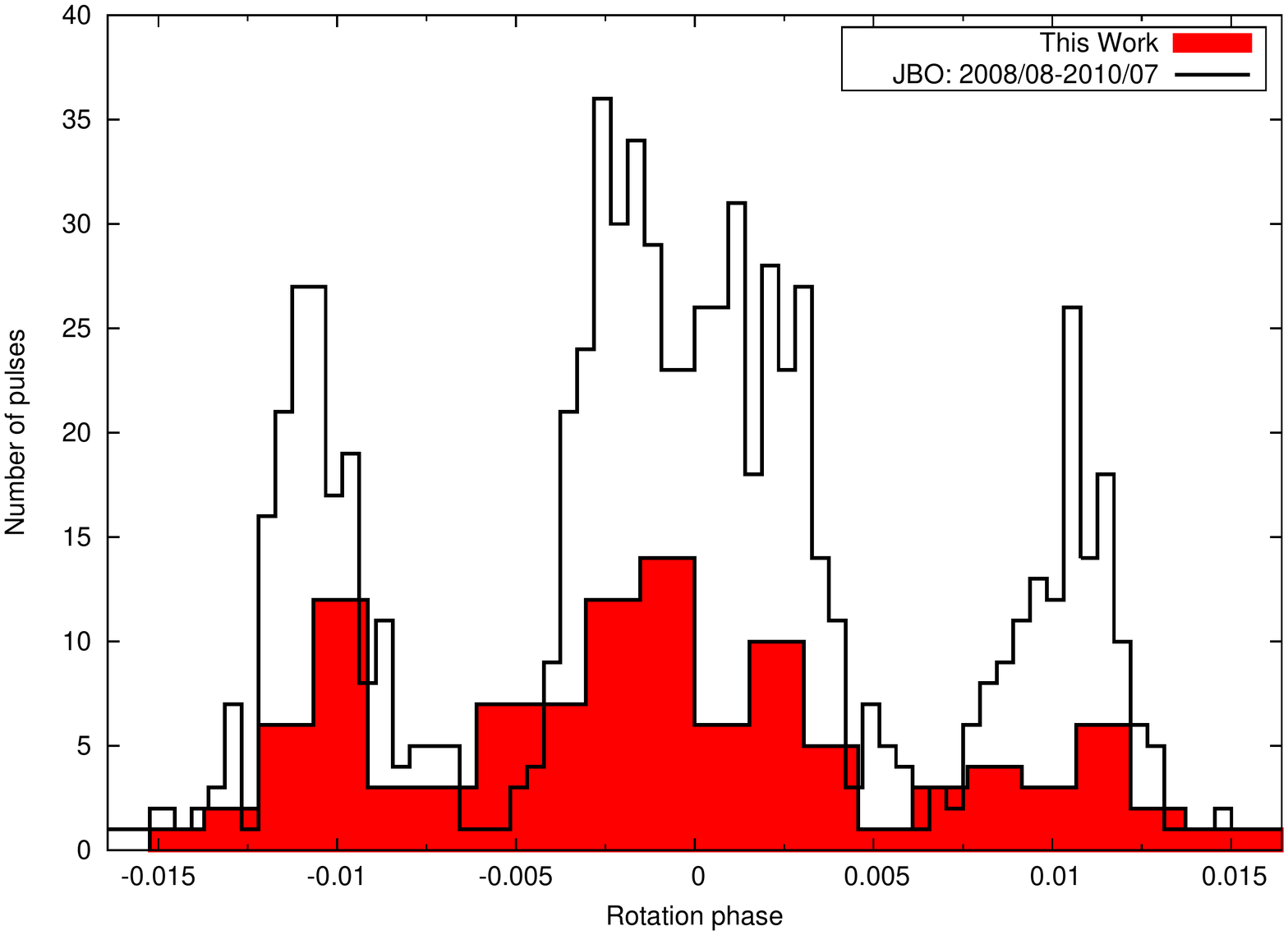}
  \caption{Left, bottom: Greyscale plot showing the
    dispersion-corrected radio pulses
    detected from J1819$-$1458 with the Lovell Telescope on 2010 June 14, where black
    indicates a higher flux density. Left, top: The mean pulse profile
    on 2010 June 14, showing the three characteristic peaks (e.g. see
    \protect{\citet{lyne09}} and~\protect{\citealt{keane10}}). Right:
    Histogram of the radio bursts from J1819$-$1458 recorded just on
    2008 August 6 and 2010 June 14 (shaded), and during the entire
    2 year period between 2008 August 6 and 2010 June 14 (unshaded).}
  \label{fig:histogram}
\end{figure*}

\begin{figure*}
  \centering
  \includegraphics[trim = 40mm 20mm 50mm 40mm, clip, scale=0.44,angle=0]{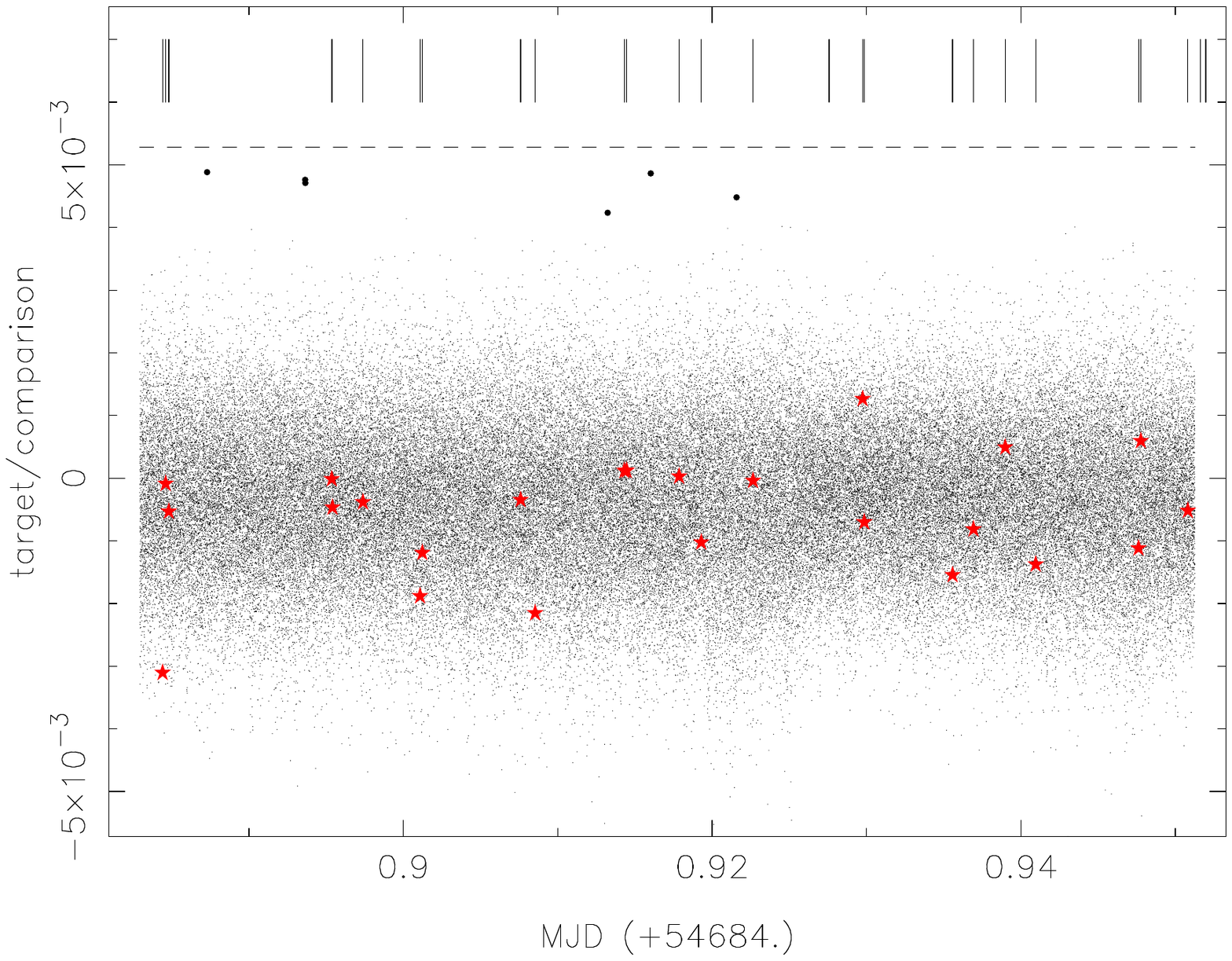}~\includegraphics[trim = 40mm 20mm 50mm 40mm, clip, scale=0.44,angle=0]{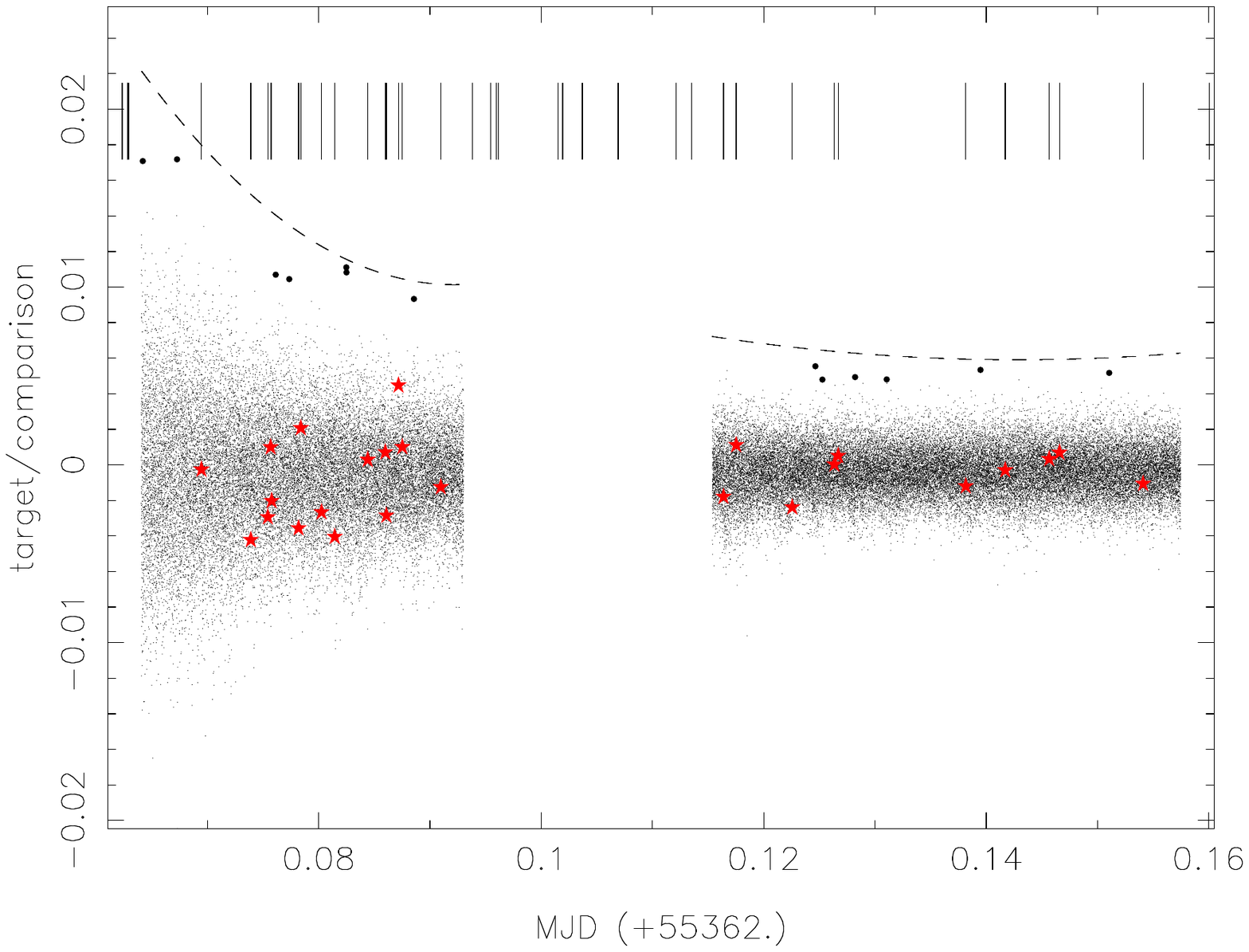}
  \caption{Left: WHT+ULTRACAM $i'$-band light curve of J1819$-$1458,
    showing all 112\,588 points observed on 2008 August 6. The dashed
    line shows the level above which points would deviate by more than
    +5$\sigma$ from the mean. The vertical bars at the top show the
    times of the dispersion-corrected radio bursts recorded by JBO. The ULTRACAM frames
    containing radio bursts are marked by the stars. For clarity, the frames lying
    above +4$\sigma$ from the mean are plotted as larger points. Right: As
    for the left, but showing the 68\,274-point $i'$-band light curve
    obtained with the NTT+ULTRACAM on 2010 June 14.  Note the
    difference in the scale of the ordinate. The increased
    scatter at the start of the run was due to poor seeing. To plot
    the dashed +5$\sigma$ curve in this case, the standard deviation
    was calculated for groups of 100 points and the result fitted with
    a polynomial. The gap in the centre of the run was due to a GRB
    override observation with ULTRACAM.}
  \label{fig:light_curve}
\end{figure*}

\section{Results}
\label{sec:results}

\subsection*{Radio observations}

The radio pulses detected from J1819$-$1458 are known to arrive
preferentially at three distinct rotation phases~\citep{lyne09}.  The
left-hand panel of Fig.~\ref{fig:histogram} shows a grey-scale
intensity plot of the individual pulses detected during the 2010 June
14 observation with the Lovell Telescope, as well as the combined
profile from adding these pulses together. The three `sub-pulses' are
clearly visible. The right-hand panel of Fig.~\ref{fig:histogram}
shows a histogram of pulse arrival times in rotation phase with
respect to the long-term radio-derived ephemeris at
JBO~\citep{lyne09}. This is essentially a probability distribution in
rotation phase for the radio pulses. The unshaded histogram denotes
all pulses detected, as part of our regular timing
observations~\citep{lyne09}, in the time interval between the optical
observations in 2008 and in 2010. The shaded histogram shows the
corresponding distribution for the pulses detecting during the optical
observations. The two histograms are similar, implying that the pulses
detected during the simultaneous observations were typical and
J1819$-$1458 seems to have been no more nor less `active' in the radio
than at other times.

\subsection*{Optical observations}
The sum of the 24 and 25 ULTRACAM $i'$-band frames containing radio
bursts on 2008 August 6 and 2010 June 14 are shown in the right-hand
central and lower panels of Fig.~\ref{fig:finding_charts}. For
comparison, the sum of all the ULTRACAM frames obtained on each night
are shown in the corresponding left-hand panels.

The circles plotted in Fig.~\ref{fig:finding_charts} indicate the
expected position of J1819$-$1458, with the radius equal to the error
in this position on our ULTRACAM frames. Inspecting the zoomed-in
boxes to the right of each panel reveals no visual evidence for the
RRAT in either burst (right) or persistent (left) light. Note the
significantly worse quality of the NTT data (bottom panels in
Fig.~\ref{fig:finding_charts}) compared to the WHT data (central
panels) due to the poorer seeing, which forced us to on-chip bin these
data by a factor of 2.

So far, we have implicitly assumed that the optical and radio bursts
are coincident, and that the radio bursts are equal or shorter in
duration than the putative optical bursts. This is a reasonable
assumption given, for example, the behaviour of the Crab pulsar, which
shows that the optical pulse is approximately 5 times wider than the
radio pulse and leads the radio pulse by only $\sim 200$\,$\mu$s
\citep{shearer03,slowikowska09}.  Since we know precisely when the
radio bursts occured, it is a simple matter to search for optical
bursts lagging/leading the radio bursts and/or of different widths to
the radio bursts by combining the appropriate optical frames. Hence,
as well as summing the ULTRACAM frames coincident with the radio
bursts, $n$, we also co-added the frames $n-1$, $n+1$ and
$n\pm1$. None of the resulting images show any evidence of the RRAT.

The searches described so far have relied on the ability of the eye to
identify a star in a summed image. A different approach is to inspect
the light curve obtained by extracting the counts in an aperture
centred on the position of the RRAT. Fig.~\ref{fig:light_curve} shows
the light curves obtained on each night. The signature of an optical
counterpart to the radio bursts would be a series of deviant points
lying approximately 5$\sigma$ or greater from the mean\footnote{We
  have chosen 5$\sigma$ as we have $\sim10^5$ data points and only one
  point in $\sim10^6$ would be expected to be greater than 5$\sigma$
  from the mean in a Gaussian distribution.} and aligned temporally
with the radio bursts (indicated by the vertical tick marks near the
top of Fig.~\ref{fig:light_curve}).  Given that the exposure time is
significantly longer than the radio burst duration, one would expect
only one, or at most two, points per burst.

The dashed line in Fig.~\ref{fig:light_curve} shows the +5$\sigma$
deviation level. It can be seen that there are no points lying above
this line. Moreover, the optical points coincident with the radio
bursts, marked by the stars in Fig.~\ref{fig:light_curve}, appear to
be randomly scattered about the mean level of zero.
Fig.~\ref{fig:light_curve} also shows the points lying 4$\sigma$ above
the mean: none of these are coincident with the radio bursts, none of
the images show any obvious sign of a star, and the intervals between
the 4$\sigma$ points are not related to the rotation period. The
implication of the preceding results is that we have not detected any
evidence for optical counterparts to the radio bursts from
J1819$-$1458. We also searched for periodicities in the light curves
using a Lomb-Scargle periodogram \citep{press89}.  No evidence for a
significant peak around the 4.263\,s rotation period, or any other
period, was found. As stated in Section~\ref{sec:observations}, the
dead-time of ULTRACAM during our J1819$-$1458 observations was always
an insignificant fraction of the exposure time and approximately the
same duration as the radio bursts shown in
Fig.~\ref{fig:histogram}. This makes it unlikely we missed a single
optical burst whilst ULTRACAM was reading out, let alone the expected
24--25 bursts.

It is useful to place a magnitude limit on the optical bursts from
J1819$-$1458 in order to constrain the spectral energy distribution.
From the summed images of the burst frames, we find that the RRAT
shows no evidence for optical bursts brighter than $i'=19.3$ at the
5$\sigma$ level in the deeper WHT observations. As expected, the
simultaneous optical-radio observations have enabled us to impose a
significantly deeper limit than the optical-only observations of paper
I, when we derived $i'>16.6$. The corresponding flux density limit is
$i'<70$ $\mu$Jy, where the flux has been calculated using equation 2
of \cite{fukugita96} and the effective wavelength of the observation
is 7610\AA\ (see \citealt{dhillon06}). We can now compare this to the
burst radio flux density of 3600 mJy at 1.4 GHz measured by
\cite{mclaughlin06} to deduce that the spectral slope must be steeper
than approximately $f_{\nu}\propto \nu^{-0.9}$.  Note that the X-ray
emission of J1819$-$1458, which exhibits pulsations rather than
bursts, is 10 nJy at 0.3-5 keV \citep{rea09}, and also lies very close to
the line $f_{\nu}\propto \nu^{-0.9}$. For comparison, the
radio-to-optical slope of the {\em pulsed} radiation from the Crab has
a much shallower slope of $\sim -0.2$ (measured from Fig. 9.3 of
\citealt{lyne05b}).

For completeness, we have also calculated the $i'$ magnitude limit of
J1819$-$1458 from the sum of all of the WHT observations. We derive a
magnitude limit of $i'>21.9$ at 5$\sigma$ confidence, significantly
deeper than the limit on the persistent light of $I=17.5$ derived by
\citet{reynolds06}. Of course, our limit could have been substantially
deeper had we decided not to look for bursts and instead taken just a
few long exposures.

\section{Discussion}

In paper I, we used aperture photometry centred on the X-ray position
of J1819$-$1458 to search for the RRAT. With no way of knowing in
which frames the bursts occurred, this search strategy was relatively
insensitive, as we were relying on detecting individual optical bursts
above the noise. In this paper, we present a far more sensitive
technique: simultaneous radio observations of the RRAT bursts to tag
the corresponding optical frames. These tagged frames were then
co-added to produce a summed `burst' frame. Unfortunately, even though
the resulting summed frame allowed us to probe nearly 3 magnitudes
deeper than before, we find no evidence for the RRAT to a 5$\sigma$
limit of $i'=19.3$. This limit allows us to say that the slope of the
pulsed radio-optical spectrum must be steeper than $-0.9$ and that
extrapolating this slope correctly predicts the X-ray flux. A more
detailed comparison with RRAT emission models is unjustified without
unambiguously determined fluxes in the optical and infrared.

In comparison to our other detections of pulsars with ULTRACAM,
e.g. $i'=25.3$ for AXP 1E\,1048.1--5937 \citep{dhillon09}, our limit
on J1819$-$1458 does not appear to be particularly deep. To place it
in some context, therefore, it should be noted that if we had taken a
single 1 hour exposure of the field with the WHT under identical
conditions, and assuming the object emitted 24 bursts, each of
$i'=19.3$ and 51.1\,ms duration, we would have have obtained a
signal-to-noise ratio of only $\sim 0.02\sigma$.  Using the high-speed
photometry technique described in this paper, on the other hand, we
would have detected the source at 5$\sigma$. The difference in
sensitivity between the two techniques is due to the fact that the
long exposure would be sky limited, whereas the data presented in this
paper are readout-noise limited, and the former noise source is over
twenty times larger than the latter. The only way we can now
significantly improve upon our magnitude limit with ULTRACAM is to
observe simultaneously for a longer period of time (in order to detect
and co-add more bursts) and/or use larger optical and radio telescopes
(in order to increase the number of counts detected from each burst in
the optical and to detect more bursts in the radio). The discussion
above assumes, of course, that the optical and radio light behave in a
similar manner. If, however, the optical light has only a low (or no)
pulsed fraction, then deep, long-exposure imaging might prove
fruitful, as might deeper searches for pulsed light on the proposed
rotation period of the neutron star, e.g. \citet{dhillon09}.

\section*{Acknowledgements}
ULTRACAM, VSD, TRM, CMC, SPL and PK are supported by the STFC. SPL
also acknowledges the support of an RCUK Fellowship. EFK acknowledges
the support of a Marie-Curie EST Fellowship with the FP6 Network
``ESTRELA'' under contract number MEST-CT-2005-19669. Based on
observations collected at ESO, Chile (Programme 085.D-0429) and the
ING, La Palma.

\bibliographystyle{mn2e}
\bibliography{abbrev,refs}

\end{document}